\documentclass[floats,floatfix,showpacs,amssymb,prd,twocolumn,superscriptaddress,nofootinbib,nolongbibliography,reprint]{revtex4-1}

\usepackage{amssymb,amsmath,verbatim,mathtools,needspace,enumitem,etoolbox,graphicx,physics,microtype,afterpage,xspace,tabularx,lmodern,multirow}
\usepackage{gensymb}
\usepackage[normalem]{ulem}
\usepackage[dvipsnames, usenames]{xcolor}
\definecolor{linkcolor}{rgb}{0.0,0.3,0.5}
\usepackage[unicode, colorlinks=true, linkcolor=linkcolor, citecolor=linkcolor, filecolor=linkcolor, urlcolor=linkcolor, linktocpage, breaklinks]{hyperref}
\usepackage[all]{hypcap}
\usepackage[T1]{fontenc}
\usepackage[utf8]{inputenc}
\usepackage[usenames,dvipsnames]{xcolor}
\hypersetup{colorlinks=true,citecolor=romared,linkcolor=romared,urlcolor=romared}

\definecolor{romared}{RGB}{142,0,28}

\newcommand{\be}{\begin{equation}}
\newcommand{\ee}{\end{equation}}

\def\be{\begin{equation}}
\def\ee{\end{equation}}
\newcommand{\beq}{\begin{eqnarray}}
\newcommand{\eeq}{\end{eqnarray}}

\usepackage{aas_macros}
\usepackage{makecell}
\usepackage{soul}
\usepackage{booktabs}
\usepackage{hyperref}

\newcolumntype{Y}{>{\centering\arraybackslash}X}

\begin{document}

\author{Konstantinos Kritos}
\email{kkritos1@jhu.edu}
\affiliation{Department of Physics and Astronomy, Johns Hopkins University, 3400 North Charles Street, Baltimore, Maryland 21218, USA}

\author{Emanuele Berti}
\email{berti@jhu.edu}
\affiliation{Department of Physics and Astronomy, Johns Hopkins University, 3400 North Charles Street, Baltimore, Maryland 21218, USA}

\author{Joseph Silk}
\email{silk@iap.fr}
\affiliation{Department of Physics and Astronomy, Johns Hopkins University, 3400 North Charles
Street, Baltimore, Maryland 21218, USA}
\affiliation{Institut d’Astrophysique de Paris, UMR 7095 CNRS and UPMC, Sorbonne Universit$\acute{e}$, F-75014 Paris, France}
\affiliation{Department of Physics, Beecroft Institute for Particle Astrophysics and Cosmology, University of Oxford, Oxford OX1 3RH, United Kingdom}

\title{Massive black hole assembly in nuclear star clusters}

\begin{abstract}
    Nuclear star clusters, which fragment into metal-poor stars {\it in situ} at the centers of protogalaxies, provide ideal environments for the formation of intermediate-mass black holes with masses $10^3$--$10^6M_\odot$. We utilize the semianalytic model implemented in \textsc{Rapster}, a public rapid cluster evolution code. We implement simple recipes for stellar collisions and gas accretion/expulsion into the code and identify the regimes where each channel contributes to the dynamical formation of intermediate-mass black holes via repeated mergers of stellar black hole seeds. We find that intermediate-mass black hole formation in gas-rich environments is almost inevitable if the initial mean density of the nuclear cluster is $>10^8M_\odot\,{\rm pc}^{-3}$. A million solar mass black hole can form within 100~Myr in the heaviest ($>10^7M_\odot$) and most compact ($<0.5~{\rm pc}$) nuclear clusters. We demonstrate that by today these resemble the observed range of nuclear clusters in dwarf galaxies and that there are potential gravitational-wave signatures of the massive black hole formation process.
\end{abstract}

\date{\today}
\maketitle

\section{Introduction}
\label{sec:Introduction}

A growing number of intermediate-mass black hole (IMBH) candidates, with masses in the range $10^3$--$10^6M_\odot$, have been identified in the centers of nearby dwarf galaxies~\cite{Reines:2013pia,Chilingarian:2018acs,2019NatAs...3..755W,2020ARA&A..58..257G}.
Often, these black holes (BHs) are surrounded by nuclear star clusters (NSCs), which are extremely crowded stellar environments~\cite{Neumayer:2020gno}.
This coexistence might suggest that NSCs have played a role in the formation of IMBHs.

Supermassive BHs, with masses of up to several billions of solar masses, are also believed to power quasars at redshift $z>6$~\cite{2003ApJ...594L..95B,2011Natur.474..616M,2015Natur.518..512W} and are found in the centers of all massive galaxies~\cite{Kormendy:2013dxa}.
The rapid emergence of these objects in the high-redshift Universe is puzzling, however, most researchers now believe that IMBH seeds may provide the missing link to the heaviest BHs observed in the Universe~\cite{Volonteri:2012tp}.
These seeds can then 
grow by a combination of clustering, merging  and gas accretion~\cite{2022MNRAS.511.2631A}.
The current challenge is then transformed into understanding how these seeds formed in the first place.

Direct collapse of pristine gas clouds at $z>10$~\cite{2010Natur.466.1082M,2022Natur.607...48L,2006MNRAS.370..289B}, gravitational runaways of stars or of BHs in the cores of star clusters~\cite{1978MNRAS.185..847B,1987ApJ...321..199Q,Miller:2012ys,Lupi:2014vza,Antonini:2016gqe,Stone:2016ryd,Sakurai:2017opi,Fragione:2020nib,Fragione:2021nhb,2021MNRAS.501.5257R,Rizzuto:2022fdp,Fragione:2023kqv}, and accretion~\cite{Panamarev:2018bwq} are a few of the channels for IMBH seed formation.
While most studies focus on each channel independently, many of these mechanisms may apply simultaneously.
This motivates us to consider IMBH growth in
NSCs that formed {\it in situ} at the centers of protogalactic disks~\cite{Devecchi:2010ps} in dwarf galaxies.
To carry out these simulations, we make use of the semianalytic model \textsc{Rapster}~\cite{Kritos:2022ggc}, which is a rapid cluster evolution public code that simulates the dynamical formation of black hole binaries.
We have implemented stellar merger and residual gas accretion/expulsion recipes into this code to consider the effects of these physical processes all at once.

In Sec.~\ref{sec:Method} we present the details of our model, including the initial conditions and the physical processes we consider.
We then identify the initial conditions for which gas accretion and stellar mergers apply (Sec.~\ref{sec:Dependence_on_initial_conditions}). In Sec.~\ref{sec:Growth_history_of_IMBHs}, we discuss the anatomy of IMBH growth through examples, and in Sec.~\ref{sec:GWs_from_eccentric_inspirals} the gravitational-wave (GW) signal associated with mergers between IMBHs and smaller BHs. In Sec.~\ref{sec:Probability_of_IMBH_formation}, we present the~results for a population of NSCs, estimate the probability for IMBH formation, and set bounds on the initial compactness of globular clusters (GCs) based on current constraints. We follow their evolution to the current epoch and compare with local observations. 
In Sec.~\ref{sec:Conclusions}, we present conclusions and directions for future work.

\section{Methods}
\label{sec:Method}

In this section, we describe the model used to perform NSC simulations (Sec.~\ref{sec:Description_of_the_model}), as well as our chosen initial conditions (Sec.~\ref{sec:Initial_conditions}). 
We then discuss the most relevant physical processes and formation channels that affect BH growth (Sec.~\ref{sec:Physical_processes}).

\subsection{The model}
\label{sec:Description_of_the_model}

Our simulations rely on the semianalytical model \textsc{Rapster}, implemented in a {\sc python} code designed to rapidly simulate the evolution of BH subsystems and the dynamical formation of binary BHs (BBHs) in the cores of star clusters~\cite{Kritos:2022ggc}. The version of the code used in this work and the relative documentation are available on {\sc github}~\cite{Rapster}.
BBH formation channels include GW captures during BH-BH hyperbolic encounters, and binary formation by the interaction of three BHs (the three-body binary, or 3bb, mechanism). The model also accounts for the formation of BBHs through a pair of exchanges from star-star to BH-star to BBHs, as well as binary-binary interactions and the formation of hierarchical triples.
It makes use of \textsc{Precession}~\cite{Gerosa:2016sys} to compute the remnant properties of BBH mergers, such as remnant mass, spin and GW recoil, while all BH spin directions are assumed to be isotropic.
The minimum and maximum simulation time steps are set to 0.1~Myr and 50~Myr, respectively.

A detailed description of the model and its assumptions can be found in Sec.~2 of Ref.~\cite{Kritos:2022ggc}.
In Sec.~\ref{sec:Physical_processes}, we elaborate on the prescriptions of those physical processes that have been added for this study: runaway stellar collisions and gas accretion/expulsion.
In our study of stellar systems in the centers of galaxies we have switched off any effects of dynamical friction and galactic tides.

\subsection{Initial conditions}
\label{sec:Initial_conditions}

We consider giant molecular clouds in the central regions of protogalactic disks within dark matter halos enriched with heavier elements from the first stars (also known as Population III stars~\cite{Bromm:2013iya}). The initial mass $M_{\rm cl,0}$ fragments into stars with metallicity above some critical threshold $Z\gtrsim10^{-4}$ (referred to as Population II stars)~\cite{Bromm:2001md,Santoro:2005tv}.
We set the metallicity to $Z/Z_\odot=0.01$,
where $Z_\odot=1.4\%$ is the solar metallicity~\cite{2009ARA&A..47..481A}.
We do not regard Population III sites (that could occur at even higher redshift), because it has been shown by cosmological hydrodynamics simulations that these form in relatively low numbers compared to our targeted NSCs in the baryon-rich cores of dark matter haloes~\cite{Susa:2014moa,2020MNRAS.494.1871W}.

Analysis of disk instabilities in high-redshift protogalactic environments in the cores of early dark matter halos showed that fragmentation of the central regions into stars can give birth to NSCs formed {\it in situ} as massive as $10^6M_\odot$, and with half-mass radii that can be as small as $0.5$~pc~\cite{Bernadetta:2008bc,Devecchi:2010ps}.
We focus on clusters with initial total mass $10^4\le M_{\rm cl,0}/M_\odot\le10^9$ for two reasons: (i) lighter systems have a lower escape velocity, and thus do not easily retain merger products due to gravitational recoils of the resulting BH~\cite{Lousto:2009mf,Kesden:2010ji,Berti:2012zp};
(ii) the cores of more massive stellar environments take longer to collapse and virialize, since they evolve more slowly (e.g. the NSC of the Milky Way has a relaxation timescale of $\simeq11$~Gyr~\cite{Panamarev:2018bwq}) so that the BH subsystem does not form rapidly enough to provide the massive seeds for SMBH growth by $z\approx6$.
We assume the initial half-mass radius of the NSC, which controls the compactness and central density, to be in the range $r_{\rm h,0}/{\rm pc}\in[0.1,3.0]$.

We parametrize the fraction of the gas cloud that forms stars by specifying the star formation efficiency, $\varepsilon$.
We take $\varepsilon\in\{3\%,10\%,30\%\}$, values which are consistent with observations of present-day star formation efficiencies~\cite{Lada:2003ss}.
The total stellar mass is given by $\varepsilon M_{\rm cl,0}$, whereas $(1-\varepsilon)M_{\rm cl,0}$ represents the residual gas that did not form stars.
To complete the description of our initial conditions, we assume that the NSC fragmented at redshift $z=13$, 
the initial binary star fraction is set to 10\%, and we take a Plummer profile for the distribution of stars.
In particular, the initial central stellar density would be given by $3\varepsilon M_{\rm cl,0}/(4\pi (r_{\rm h,0}/1.3)^3)$, and it evolves self-similarly.

The choice of NSCs forming at redshift $z=13$ is somewhat arbitrary and coincides with the era of the formation of the first galaxies. It also allows for evolving NSCs for about a Hubble time and corresponds to the distance range below which the space-borne GW detector LISA is expected to be sensitive. Moreover, the observation of SMBHs in quasars at $z\sim7$ motivated the above choice, which is necessary in order to explain the formation of such massive BHs at high redshift.

\subsection{Physical processes}
\label{sec:Physical_processes}

\subsubsection{Core collapse}

Star clusters evolve internally via two-body relaxation. The timescale over which energy is shared among the stars as the system attempts to establish thermal equilibrium is given by the half-mass relaxation timescale, and it is expressed as follows~(see e.g. Ref.~\cite{1987degc.book.....S}, page~40):
\begin{align}
    \tau_{\rm rh}&=\left({r_{\rm h}\over GM_{\rm cl}}\right)^{1\over2}{N\over8\ln\Lambda}\nonumber\\&\simeq843~{\rm Myr}\cdot\varepsilon\left({r_{\rm h}\over1~\rm pc}\right)^{3\over2}\left({M_{\rm cl}\over10^7M_\odot}\right)^{1\over2}{0.7M_\odot\over\overline{m}}{10\over\ln\Lambda},
    \label{eq:half_mass_relaxation_timescale}
\end{align}
where $N=\varepsilon M_{\rm cl}/\overline{m}$ is the number of stars in the cluster, and $\overline{m}$ is the average mass $\simeq0.7M_\odot$ for a Kroupa (2001) initial mass function (IMF)~\cite{Kroupa:2000iv}.
Here, we emphasize that $M_{\rm cl}$ denotes the total mass in stars and gas and $r_{\rm h}$ is the half-mass radius. 
The Coulomb logarithm is approximated by $\Lambda\simeq0.1N$.
To account for the effect of a population of BHs in the late evolution of the cluster, we divide the relaxation timescale by the multimass ``psi'' $\psi$ factor using Eq.~(12) from~\cite{Antonini:2019ulv}, and write $\tau_{\rm rh}\to\tau_{\rm rh}/\psi$.

The ultimate fate of self-gravitating systems is to evolve towards a state of core collapse with expanding outer layers, a phenomenon known as {\it gravothermal catastrophe}~\cite{1968MNRAS.138..495L}.
In systems with a broad mass spectrum, the timescale for core collapse is~\cite{PortegiesZwart:2002iks}
\begin{align}
    \tau_{\rm cc}\simeq0.20\tau_{rh,0},
    \label{eq:core_collapse_timescale}
\end{align}
that is, a fraction of the initial half-mass relaxation timescale.
Typically, the formation of hard binaries in the center of the system provides the energy source required to halt the collapse of the system into a singularity~\cite{1975IAUS...69..133H}.
Moreover, the heaviest objects in the cluster (BHs in particular) tend to sink and settle into the core of the system, a process known as {\it mass segregation}~\cite{1969ApJ...158L.139S}.
Nevertheless, complete energy equipartition is rarely attained in collisional stellar systems~\cite{Trenti:2013ety}.
A BH subsystem then forms, which persists in the core for up to a Hubble time~\cite{Breen:2013vla,Morscher:2014doa}.

In \textsc{Rapster}, whenever an interaction occurs among BHs, we draw masses from the list of available BHs in the system according to the dependence of the reaction rate on mass. 
This choice favors interactions of the heaviest BHs in our simulations.

\subsubsection{Runaway stellar collisions}
\label{sec:Runaway_stellar_collisions}

When the core of a star cluster collapses, the central density and reaction rates rise dramatically.
If the core collapse timescale $\tau_{\rm cc}$ occurs faster than massive stars evolve to compact objects, those stars will condense into the central region of the system and, due to finite-size effects, physical stellar collisions will become inevitable~\cite{PortegiesZwart:2002iks}.
The lifetime of the most massive stars is below 4~Myr (see Table~I in~\cite{2002RvMP...74.1015W}), with some dependence on metallicity~\cite{Hurley:2000pk}.
Furthermore, we extrapolate the stellar IMF to $150M_\odot$, which is believed to be the maximum stellar initial mass based on observations of stellar masses in the Arches cluster~\cite{Figer:2005gr}.
Since we simulate massive low-metallicity giant molecular clouds which fragment into millions of stars, we almost always find at least one star with a mass above $100M_\odot$ for which the lifetime is $\approx3$~Myr.
Based on these considerations, we account for the effect of runaway stellar collisions when $\tau_{\rm cc}<3$~Myr, i.e., when
\begin{align}
    {M_{\rm cl,0}\over10^7M_\odot}\lesssim3.2\times10^{-4}{1\over\varepsilon^2}\left({r_{\rm h,0}\over\rm 1~pc}\right)^{-3}.
    \label{eq:Stellar_collisions_condition}
\end{align}
In this case, massive stars will start coalescing with each other to form a runaway massive star, which later collapses into an IMBH within a few Myr.
If the condition~\eqref{eq:Stellar_collisions_condition} is not met, then the most massive stars will form BHs on a timescale smaller than the core collapses, at which point the core will be BH dominated.

To account for runaway stellar collisions we implement a simple model into \textsc{Rapster} based on Refs.~\cite{PortegiesZwart:2002iks,Glebbeek:2008in,1997MNRAS.291..732T}.
According to simulations performed by Portegies Zwart {\it et al.} (1999)~\cite{PortegiesZwart:1998nb}, it is typically the heavier stars that begin the runaway process.
We assume that once the core collapses at $t=\tau_{\rm cc}$, the first stellar collision occurs, and that it involves the heavier star (the seed) in the cluster~\cite{PortegiesZwart:2002iks}.
As in Ref.~\cite{PortegiesZwart:2002iks}, the minimum stellar mass $m_f(t)$ that sinks into the core at time $t$ in the simulation is inversely proportional to $t$ [see Eq.~(9) from that reference], because the dynamical friction timescale depends on the mass $m$ as a fraction $\sim\overline{m}/m$ of the half-mass relaxation timescale.
To carry out the stellar collision at time $t$, we draw the mass $m_s\ge m_f(t)$ of the secondary from the list of stellar masses as sampled from the IMF.
In order to favor the collision of heavier stars with the runaway, we draw $m_s$ according to the distribution $p(m_s)\propto m_s^{3/2}$, based on the mass dependence of the reaction rate.
The timescale between successive collisions of the runaway with other stars is computed as follows [see Eq.~(17) from~\cite{PortegiesZwart:2002iks}]:
\begin{align}
    \tau_{\rm coll}\simeq0.23~{\rm Myr}{\tau_{\rm rh}\over{500~\rm Myr}}{10^7\over N}.
    \label{eq:collision_timescale}
\end{align}

After the collision, the two main sequence stars fuse together into a single new main sequence star.
The mass $m_r'$ of the collision product is $m_r'=(1-\phi)(m_r+m_s)$, where $m_r$ is the mass of the runaway star and $\phi$ is the fraction of mass lost during the collision (see the fit given in Ref.~\cite{Glebbeek:2008in}).
Since $m_s\ll m_r$, often this equation reduces to $m_r'\simeq m_r+m_s$, known as the sticky-sphere approximation~\cite{Kremer:2020wtp}.

Typically, the runaway star (which is the primary) is the most evolved of the two stars involved in the merger.
Therefore, the secondary supplies fresh hydrogen fuel and extends the lifetime of the runaway star, provided efficient H/He mixing is achievable~\cite{Lombardi:1995iu}.
This rejuvenation process occurs only when both stars are burning hydrogen. We compute the age of the collision product as~\cite{1997MNRAS.291..732T,Hurley:2002rf}
\begin{align}
    \tau_r' = f_{\rm rej}{\tau_{_{\rm MS}}(m_r')\over m_r'}\left[{m_r\tau_r\over\tau_{_{\rm MS}}(m_r)} + {m_s\tau_s\over\tau_{_{\rm MS}}(m_r)}\right],
\end{align}
where $\tau_r$ and $\tau_s$ are the stellar ages of the runaway and the secondary before the collision, respectively, and $\tau_{_{\rm MS}}(m)$ is the lifetime of a star with mass $m$ on the main sequence. We set $f_{\rm rej}$, which quantifies the amount of rejuvenation experienced by the stellar collision remnant via hydrogen mixing, equal to 1.0, as in Ref.~\cite{Kremer:2020wtp}.

We update the simulation time at every collision and set the time step equal to $\tau_{\rm coll}$ [cf.~Eq.~\eqref{eq:collision_timescale}].
Over time, lighter stars penetrate the core of the cluster, and it becomes more likely that the runaway swallows smaller stars.
Thus, the growth of the runaway saturates. 
The runaway persists as long as $\tau_{\rm coll}<\tau_{_{\rm MS}}(m_r)-\tau_r$, i.e., when the remaining lifetime of the massive star is longer than the collision timescale, or when 50\% of massive stars have collapsed into BHs.
The latter condition is a conservative choice we make as a termination criterion for the stellar runaway growth.

We neglect the effect of stellar winds on the runaway. Stellar winds are highly uncertain for massive stars. Generally speaking, low-metallicity stars (which we are dealing with) are expected to have smaller opacity due to the reduction of metals in the stellar atmosphere~\cite{Vink:2001cg,2022arXiv220308218B}, and it is possible that winds would have a small effect on the growth of the IMBH.
See for example Fig.~2 in Ref.~\cite{Mapelli:2018uds}, where a massive metal-poor star loses only a small fraction of its original mass due to winds.
The same however cannot be said for higher metallicity environments, where it has been demonstrated that stellar winds will likely limit the growth of the runaway significantly~\cite{Mapelli:2016vca,2009A&A...497..255G}.
However, we are not simulating these solar-metallicity environments.

\begin{figure}
    \centering
    \includegraphics[width=0.5\textwidth]{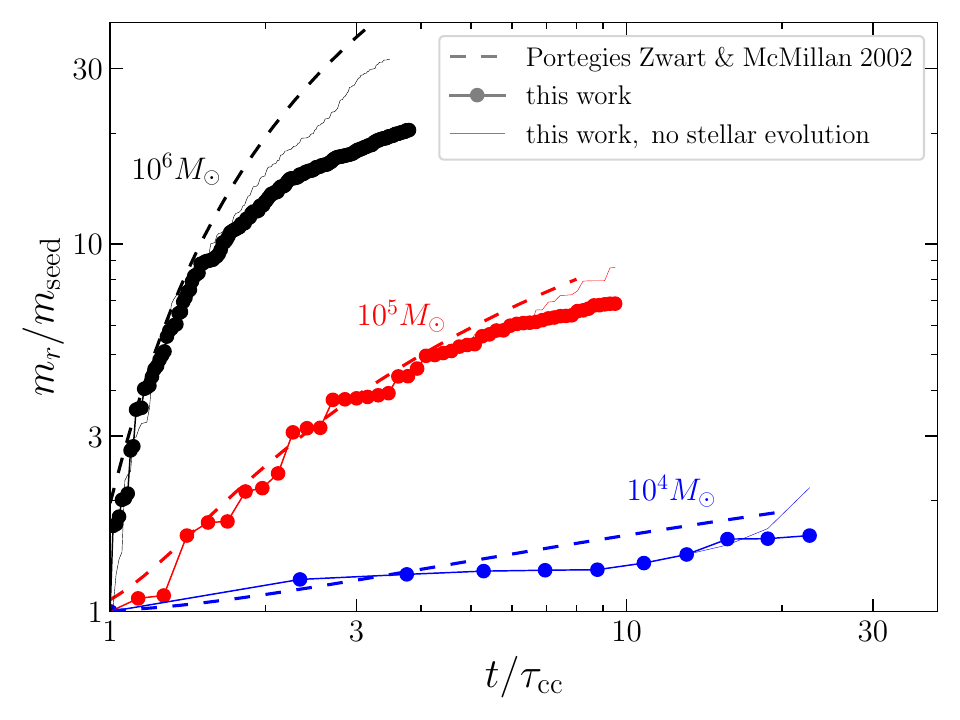}
    \caption{Comparison of our stellar collision model with the analytical prescription of Ref.~\cite{PortegiesZwart:2002iks}. We plot the mass $m_r$ of the runaway star normalized to the seed mass $m_{\rm seed}$ as it grows in time (where time $t$ is measured in units of the core collapse timescale $\tau_{\rm cc}$ for each cluster) for three values of the cluster mass, as indicated in the figure. Each filled dot corresponds to a stellar collision. The initial half-mass radius is $r_{\rm h,0}=0.1$~pc for all three clusters. The thin solid lines show the seed growth when we ignore the effect of stellar evolution.}
    \label{fig:stellar_collisions}
\end{figure}

In Fig.~\ref{fig:stellar_collisions} we show that our stellar collision model is in good agreement with the analytical prescription of Ref.~\cite{PortegiesZwart:2002iks}, as it should be (after all, our collision rate follows that reference). However, we predict slightly smaller masses near the end of the collisional runaway, especially in massive clusters.
We have checked that during the simulation some massive stars that do not undergo coalescence evolve into BHs in real time as the runaway coalescence proceeds. Thus, the mass of the runaway star saturates, as it is only fed by lighter stars. This is more realistic than the analytical formula presented in~\cite{PortegiesZwart:2002iks}, which does not take into account stellar evolution. We have verified this hypothesis by turning off stellar evolution, and indeed we get a better agreement with Ref.~\cite{PortegiesZwart:2002iks}, as shown by the thin solid lines in Fig.~\ref{fig:stellar_collisions}. In our main simulations, we do allow for the depletion of heavier-mass runaway stars due to stellar evolution.

\subsubsection{Residual gas accretion and removal}
\label{sec:Residual_gas_accretion_and_removal}

Only a fraction $\varepsilon$ of the original giant molecular cloud turns into stars, and for the first Myr, the star cluster is embedded into a gas cloud~\cite{Lada:2003ss}.
The remaining gas within the cluster does not settle into the core forever, but it is expelled from the system altogether due to stellar feedback (radiation and winds from OB stars)~\cite{2008IAUS..246...36B}.
The expulsion timescale of this residual gas due to feedback from OB stars depends on the depth of the potential well of the cluster, and is given by~\cite{Baumgardt:2007jb}
\begin{align}
    \tau_{\rm ge}^{OB}\simeq0.71~{\rm Myr}{1-\varepsilon\over\varepsilon}{M_{\rm cl}\over10^7M_\odot}{1\,{\rm pc}\over r_{\rm h}}.
    \label{eq:residual_gas_expulsion_timescale_OB}
\end{align}

It is usually assumed that any remaining gas is removed from the cluster on timescales smaller than that of stellar evolution, thus the impact of residual gas accretion onto BHs is ignored in most studies.
However it could have a significant effect if the gas expulsion timescale $\tau_{\rm ge}$ is longer than the formation of the compact BH subsystem.
Therefore, BHs accrete from the residual gas until its complete dispersion as long as $\tau_{\rm ge}>\tau_{\rm cc}$.
Evidently, massive compact clusters, with deep potential wells, expel their gas component on a longer timescale, thus allowing for stellar BHs (SBHs), with masses up to $\sim100M_\odot$, to feed from gas.
Moreover, the smaller $\varepsilon$, the larger the amount of residual gas and the smaller the core collapse timescale, since the latter scales with the number of stars: cf.~Eq.~\eqref{eq:half_mass_relaxation_timescale}.

Individual supernova (SN) explosions release large amounts of energy, typically of order $10^{51}$~erg.
The total amount of energy produced by all SNe altogether, $E_{\rm SN}$, is given by Eq.~(15) from~\cite{Baumgardt:2007jb}.
We denote by $E_{\rm gas}\simeq0.4(1-\varepsilon)GM_{\rm cl}^2/r_{\rm h}$ the binding energy of the residual gas cloud in the cluster.
If $E_{\rm SN}>E_{\rm gas}$, feedback from supernovae is capable of removing the gas from the cluster on a timescale of $t_{ge}^{\rm SN}=3$~Myr~\cite{Baumgardt:2007jb}.
However, if  $E_{\rm SN}<E_{\rm gas}$, feedback from SNe cannot unbind the gas cloud from the cluster, and its removal proceeds by stellar winds and radiation on a longer timescale.
In that case, formally, we have $\tau_{\rm ge}^{\rm SN}=\infty$.

Finally, the gas cannot disperse faster than the dynamical time, $r_{\rm h}/c_s$, which sets the lower bound for the gas expulsion timescale $\tau_{\rm ge}$.
Putting all the pieces together, and assuming that the total power released is the sum of the power produced by each different feedback mechanism, we compute $\tau_{\rm ge}$ as
\begin{align}
    \tau_{\rm ge}=\max\left\{{r_{\rm h}\over c_s}, \min\left[ \tau_{\rm ge}^{OB}, \tau_{\rm ge}^{SN} \right]\right\}.
    \label{eq:gas_expulsion_time}
\end{align}

We assume that the gas (primarily composed of hydrogen) disperses and its mass evolves exponentially with time~\cite{Baumgardt:2007av}, but unlike that reference, we do not consider an initial time delay in gas expulsion:
\begin{align}
    M_{\rm gas}(t)=(1-\varepsilon)M_{\rm cl,0}e^{-t/\tau_{\rm ge}}.
    \label{eq:gas_mass_evolution}
\end{align}
The sound speed in the gas is set to $c_s=10~\rm km/s$~\cite{2009ApJ...703.1352K}.
Assuming the gas is homogeneous and follows the Plummer profile, its mass density in the core is given by $\rho_{\rm gas}(t)\simeq0.53M_{\rm gas}(t)/r_{h,\rm gas}(t)^3$.
The half-mass radius of gas evolves according to Eq.~(3) from Ref.~\cite{Kroupa:2000fr}, and at time $t=0$ we set $r_{h,\rm gas}(t=0)=r_{\rm h,0}$ as the initial condition.

We grow the mass of all in-cluster BHs (both single and in binaries) by applying the Bondi accretion rate formula for $dm_{\rm BH}/dt$ capped by the Eddington limit, using a time-dependent expression for the density of gas, $\rho_{\rm gas}(t)$, as above. In particular, at every step of the simulation we increase the mass of every BH in the cluster by an amount $\Delta m_{\rm BH}=({dm_{\rm BH}/dt})\Delta t$, where $\Delta t$ is the current time step of the simulation (note that this is an adaptive step).
For simplicity, we do not implement spin-up prescriptions for the accreting BHs.

\subsubsection{Black hole mergers}

Once the residual gas disperses away from the cluster, the BH subsystem becomes dry.
The massive BH can now continue growing---or emerge, if it did not form to begin with---by repeated mergers of stellar-mass BHs, remnants of stellar evolution.
Below, we discuss the limitation of hierarchical BH growth due to GW kicks imparted to the merger remnant (which can eject it from its host environment), and then describe the dynamical formation scenarios that contribute to the formation channels of BBHs in our model.
We start with our assumptions on the initial BH mass spectrum.

\begin{figure*}[t]
    \centering
    \includegraphics[width=\textwidth]{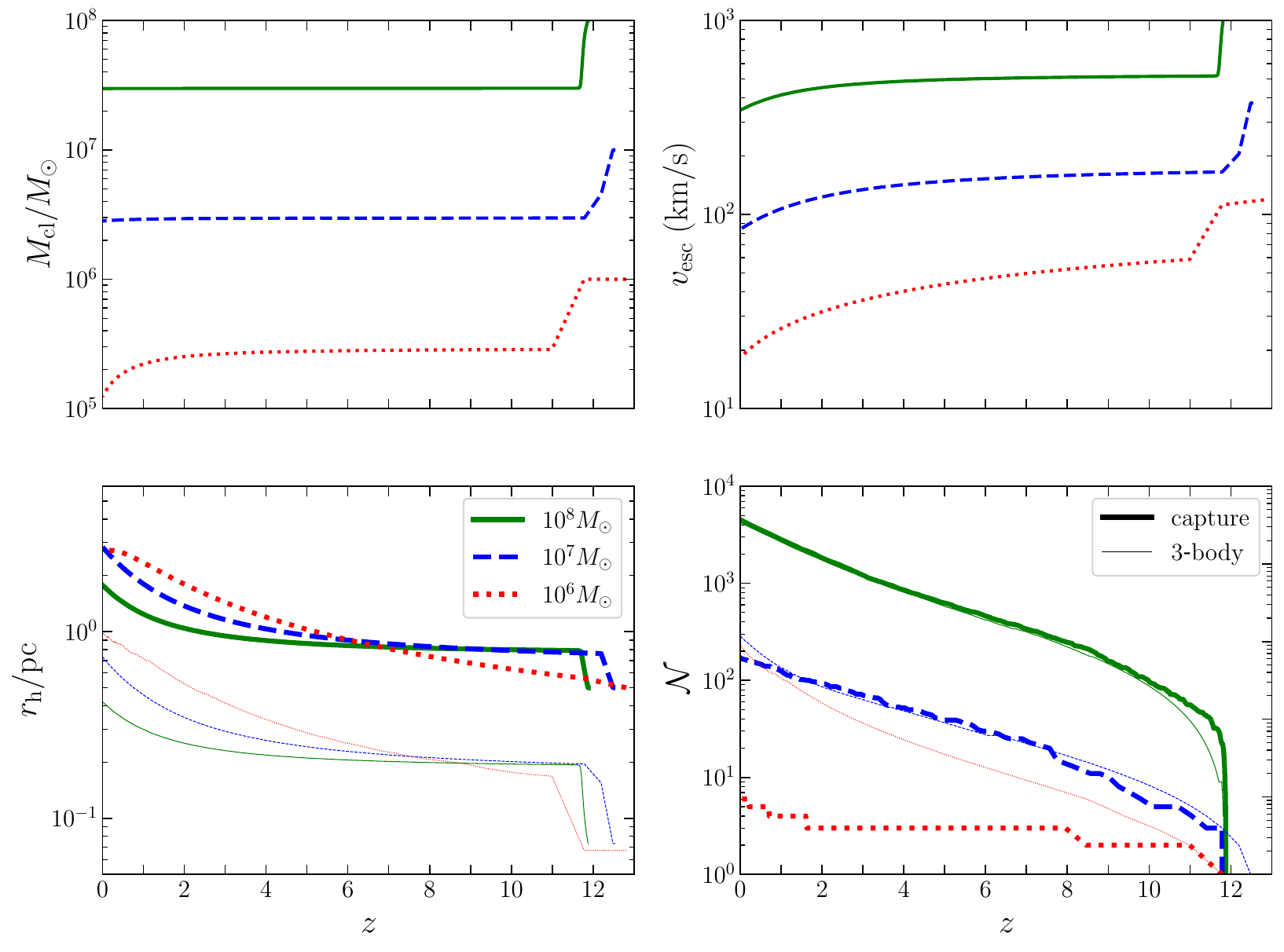}
    \caption{
    Redshift evolution of cluster mass (upper left), half-mass radius (lower left), escape velocity (upper right), and cumulative number of captures and three-body binary formation (lower right).
    We show the evolution of models with $r_{\rm h,0}=0.5$~pc and initial cluster masses $10^8M_\odot$ (green), $10^7M_\odot$ (blue), and $10^6M_\odot$ (red), while $\varepsilon=30\%$ and $Z=0.01Z_\odot$.
    The thin lines in the lower-left panel correspond to the evolution of the segregation radius of BHs for the three models considered in this figure.}
    \label{fig:cl_evolution}
\end{figure*}

\noindent
\paragraph{Natal BH masses.---}

We compute the mass of SBHs, remnants of stellar evolution, using the remnant-mass prescription of Ref.~\cite{Fryer:2011cx} with pulsational pair-instability and pair-instability supernova conditions from Ref.~\cite{Belczynski:2016jno}. We make use of the ``delayed'' supernova engine (the ``rapid'' model differs from the delayed model only for BH masses below $20M_\odot$, but these are unlikely to be relevant for dynamical environments, because the most massive BHs are the ones that segregate strongly to the center and react more often).
In order to evolve single stars, we make use of the {\sc SSE} subroutine from the publicly available, updated {\sc BSE} package~\cite{Banerjee:2019jjs}.
The remnant-mass prescription is extrapolated to very massive stars, and we apply Eq.~(7) from Ref.~\cite{Fryer:2011cx} to predict the mass of the IMBH that forms from the process of runaway collisions.
In particular, for stars more massive than a few hundred $M_\odot$, the mass of the IMBH remnant is given by
\begin{align}
    {M_{\rm IMBH}}= \left(1-1.3\sqrt{Z\over Z_\odot}\right)m_\star + 18.35M_\odot\sqrt{Z\over Z_\odot},
    \label{eq:DC_IMBH}
\end{align}
where $m_\star$ and $Z$ are the mass and absolute metallicity of the progenitor star, respectively.
Moreover, we assume BHs are born nonspinning. Second- and higher-generation BHs obtain a nonzero spin by inheriting the orbital angular momentum of the progenitor BBH, as calculated with \textsc{Precession}~\cite{Gerosa:2016sys}.
\\

\noindent
\paragraph{GW kick and BH retention.---}
\label{sec:GW_kick_and_BH_retention}

When two BHs coalesce, the merger remnant obtains a significant recoil velocity due to asymmetric GW emission~\cite{1983MNRAS.203.1049F}.
For instance, the GW kick is about 175~km/s when the mass ratio is $q\approx0.36$ and the BHs are nonspinning~\cite{Gonzalez:2006md}.
Moreover, second-generation BHs---products of first-generation BH mergers---typically have large spins ($\chi\approx0.7$~\cite{Tichy:2008du,Berti:2008af}) and the merger remnant might even obtain a superkick of thousands of km/s, with a sensitive dependence on the spin configuration~\cite{Campanelli:2007cga}.
Hierarchical mergers are limited by the escape speed of the host environment.
Therefore, BH growth via repeated mergers requires a massive dense environment for which the escape velocity is larger than a ${\rm few}\times100$~km/s~\cite{Merritt:2004xa,Antonini:2016gqe,Gerosa:2019zmo,Gerosa:2020bjb,Gerosa:2021mno}, and NSCs are exceptional examples.
In particular, the escape velocity from a cluster with total (stars + gas) mass $M_{\rm cl}$ and half-mass radius $r_{\rm h}$ is given by (Ref.~\cite{1987degc.book.....S}, page~12)
\begin{align}
    v_{\rm esc}=2\sigma\simeq262~{\rm km/s}\left({M_{\rm cl}\over10^7M_\odot}\right)^{1\over2}\left({r_{\rm h}\over1~{\rm pc}}\right)^{-{1\over2}}.
    \label{eq:escape_velocity}
\end{align}

The properties of a merger remnant (mass, spin, and GW kick) depend crucially on the mass ratio of the progenitor BBH and BH spins~\cite{Lousto:2009mf}.
Those are computed in \textsc{Rapster} implementing the \textsc{Precession} code~\cite{Gerosa:2016sys}. 
As the runaway BH grows in mass through a series of successive coalescence episodes and the mass ratio tends to zero, the remnant spin is dominated by the primary's spin, and given the randomized orientations the latter tends to decrease with time~\cite{Hughes:2002ei}.
Obviously, assuming the merger remnants are retained in the same dense environment, the mass increases at a steady rate, however the spin of the BH asymptotically approaches to zero executing a random walk.
Such an anticorrelation between the mass and spin of the runaway BH has been reported before~\cite{Fragione:2021nhb,Zevin:2022bfa,Berti:2008af}.
\\

\noindent
\paragraph{Dynamical BBH formation channels.---} 

In star clusters, BBHs form primarily via two-body and three-body interactions. For a description of our treatment for these processes in \textsc{Rapster}, see Sec.~II of Ref.~\cite{Kritos:2022ggc}.
These include GW captures, which occur during close BHBH hyperbolic encounters and the interaction of three single BHs in the core, with the third BH carrying away the excess kinetic energy required to be converted into the internal degree of freedom (binding energy) of the induced BBH.
This is also known as the three-body binary (or 3bb) channel.

The formation of BBHs via a pair of exchanges from binary stars, process of the form star-star$\to$BH-star$\to$BH-BH is also implemented in \textsc{Rapster}; however, this exchange channel is quenched for very massive dense clusters, because in those environments the hard-soft boundary occurs at very small values of the semimajor axis (since the velocity dispersion is very high)~\cite{Sesana:2006xw}:
\begin{align}
    a_{\rm h}={G\overline{m}\over4\sigma^2}\simeq3.3R_\odot{\overline{m}\over 0.7M_\odot}\left({\sigma\over100~{\rm km/s}}\right)^{-2}.
    \label{eq:hardness_sma}
\end{align}
Thus, most hard binary stars are near-contact binaries, and the binary cross section is small enough to make the exchange channel inefficient for massive clusters. In particular, we do not observe exchanged pairs in our models of massive clusters.

The strong dependence of the 3bb rate on the velocity dispersion $\sigma$ (it scales as $\sigma^{-9}$) makes the 3bb channel rare in those NSCs. However, the absence of hard BBHs in the core leads to its collapse, and the dramatic isothermal increase in the central density overcomes the $\sigma^{-9}$ dependence~\cite{Lee:1994nq}.
Hence, a large number of three-body hard BBHs still form in those massive clusters able to stop complete core collapse.
On the other hand, BBHs that merge on a time smaller than the interaction timescale (such as GW captures) cannot heat up the cluster~\cite{1993ApJ...408..496M}.

\subsubsection{Cluster evolution}

Our time evolution simulations begin once the core has collapsed, at time $t=\tau_{\rm cc}$: cf.~Eq.~\eqref{eq:core_collapse_timescale}.
We track the evolution of the cluster as an isolated system, by modeling mass loss due to gas expulsion and star loss from two-body relaxation.
We have modified the \textsc{Rapster} model to ignore the effect of tidal stripping mass loss from the host galaxy, since we have assumed the nuclear cluster rests in the center of the galactic potential well.
Hard binaries in the core generate energy that heats up the whole cluster on the half-mass relaxation timescale $\tau_{\rm rh}$ of Eq.~\eqref{eq:half_mass_relaxation_timescale}, and causes the cluster to slowly expand.
In the evolution of half-mass radius, we also account for the expansion due to gas dispersion by adding an adiabatic term.
In summary, the semianalytic differential equations that govern the evolution of the cluster mass $M_{\rm cl}=M_{\rm gas}+M_\star$ and half-mass radius $r_{\rm h}$ are given by~\cite{Gnedin:2013cda,Antonini:2019ulv}
\begin{subequations}
\begin{align}
    {dM_{\rm cl}\over dt}&={dM_{\rm gas}\over dt} - 2.5\xi_e {M_\star\over \tau_{\rm rh}}\label{eq:mass_evolution},\\
    {dr_{\rm h}\over dt}&=\zeta{r_{\rm h}\over \tau_{\rm rh}} + 2{r_{\rm h}\over M_{\rm cl}}{dM_{\rm cl}\over dt} - {r_{\rm h}\over M_{\rm cl}}{dM_{\rm gas}\over dt},\label{eq:radius_evolution}
\end{align}
\end{subequations}
where $\zeta=0.0926$ and $\xi_e=0.0074$.
The first term in Eq.~\eqref{eq:mass_evolution} is calculated by differentiating Eq.~\eqref{eq:gas_mass_evolution}.
Notice that in our model, $M_\star$ accounts for both the stellar and BH components.
We solve these differential equations numerically as the simulation proceeds, with the time step $dt_1$ set by the algorithm of the code (see Sec.~II.E.2 of Ref.~\cite{Kritos:2022ggc}).

The redshift evolutions of cluster mass, half-mass radius, escape velocity and GW capture rate are shown in Fig.~\ref{fig:cl_evolution} for clusters with $r_{\rm h,0}=0.1~\rm pc$.
Since $\varepsilon=0.3$ for these models, the mass of the clusters drops to $30\%$ of its initial value due to rapid gas expulsion in the initial phase on short timescales ($\sim$Myr).
Subsequent evolution proceeds via two-body relaxation on longer timescales ($\sim$Gyr).
The half-mass radius expands from 0.5~pc to a few~pc within a Hubble time (Fig.~\ref{fig:cl_evolution}, lower left panel), which is typical of the effective radii of present-day observed NSCs at $\sim3~{\rm pc}$~\cite{2016MNRAS.457.2122G}.
The BHs settle within a smaller volume, with typical segregation radii that are on average a factor of $\sim3$ smaller than the cluster's half-mass radius.
The evolution of the BH segregation radii is shown by thin lines in the lower left panel of Fig.~\ref{fig:cl_evolution}.
Finally, the interaction rates decline with time as a consequence of the expansion of the BH subcluster, and its present-day value depends on the cluster mass.
For properties similar to the NSC that reside in the center of our Milky Way ($M_{\rm cl}\simeq2\times10^7M_\odot$ and $r_{\rm h}\simeq4~{\rm pc}$)~\cite{2014CQGra..31x4007S}, our simulations indicate that the capture rate in its core could be of the order of one GW capture every 10~Myr.

As evidenced by the lower-right panel of Fig.~\ref{fig:cl_evolution}, the number of captures (thick lines) becomes more significant with increasing mass when compared to the cumulative number of three-body binaries (thin lines).
Even though a large number of three-body binaries still form in heavier NSCs, a smaller fraction of them eventually merges, thus making the two-body capture a more efficient merger channel in those environments.
In the Appendix, we explore the validity of our assumption that three-body BH binaries can support the star cluster via binary-single interactions within the BH subsystem.
We show that the vast majority (if not all) of our simulations---including those we present in Fig.~\ref{fig:cl_evolution}---have conditions that allow for efficient heating of the cluster from three-body binaries.

\section{Initial conditions}
\label{sec:Dependence_on_initial_conditions}

\begin{figure*}
    \centering
    \includegraphics[width=1\textwidth]{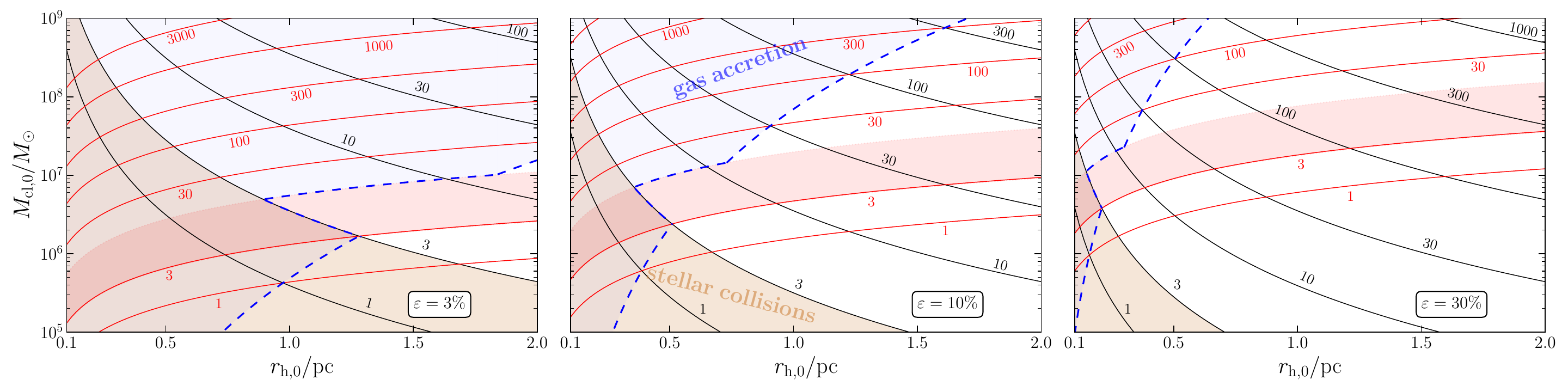}
    \caption{Core-collapse time $\tau_{\rm cc}$ (black, negative-slope contours) [cf.~Eq.~\eqref{eq:core_collapse_timescale}] and gas-expulsion timescale $\tau_{\rm ge}$ (red, positive-slope contours) [cf.~Eq.~\eqref{eq:gas_expulsion_time}] in the parameter space of initial cluster mass $M_{\rm cl,0}$ and initial half-mass radius $r_{\rm h,0}$. Contour labels refer to timescales in Myr. The blue region above and to the left of the blue-dashed curves corresponds to clusters for which $\tau_{\rm cc} < \tau_{\rm ge}$. The red stripes correspond to clusters with a gas expulsion timescale of 3~Myr, the typical lifetime of massive stars. The peru-filled regions correspond to clusters for which $\tau_{\rm cc}<3~{\rm Myr}$. The panels show the contour plots for three representative values of $\varepsilon$, the star formation efficiency.}
    \label{fig:param_sp}
\end{figure*}

Depending on the initial conditions of NSCs, mass $M_{\rm cl,0}$ and half-mass radius $r_{\rm h,0}$, the IMBH can emerge as a consequence of runaway stellar mergers and/or repeated BH mergers.
In Fig.~\ref{fig:param_sp} we identify the regions of the $r_{\rm h,0}-M_{\rm cl,0}$ parameter space where effects of stellar mergers and residual gas accretion affect IMBH growth, selecting three typical values of star formation efficiency of $\varepsilon=3\%,10\%,30\%$ (consistent with observations in the local Universe~\cite{Lada:2003ss}).

Stellar mergers are likely to trigger IMBH growth in lighter clusters, because their core collapse and relaxation timescales are small enough for massive stars to concentrate and collide before they evolve into compact objects.
In particular, runaway stellar collisions occur if the core of the cluster collapses on a timescale less than the lifetime of massive stars, which is around 3~Myr.

Residual gas, not turned into stars, may be accreted by the central IMBH that just started to grow, as well as by smaller BHs, before the gas disperses away due to stellar feedback (radiation from OB stars and supernovae).
Nevertheless, for gas to have an effect on the growing IMBH, it should be retained for tens or hundreds of Myr, and this requires very deep potential wells that prolong its presence in the cluster.
As seen from Fig.~\ref{fig:param_sp}, this is achieved for the most massive compact clusters, as the binding energy of the gas cloud is proportional to $(1-\varepsilon)M_{\rm cl}^2/r_{\rm h}$~\cite{Baumgardt:2007jb}.
In those cases---with initial conditions that lie to the left of the blue-dashed line in Fig.~\ref{fig:param_sp}---the gas expulsion timescale is much larger than the time necessary for the formation of the central BH subcluster.

\section{Growth history of intermediate-mass black holes}
\label{sec:Growth_history_of_IMBHs}

\begin{figure}
    \centering
    \includegraphics[width=0.5\textwidth]{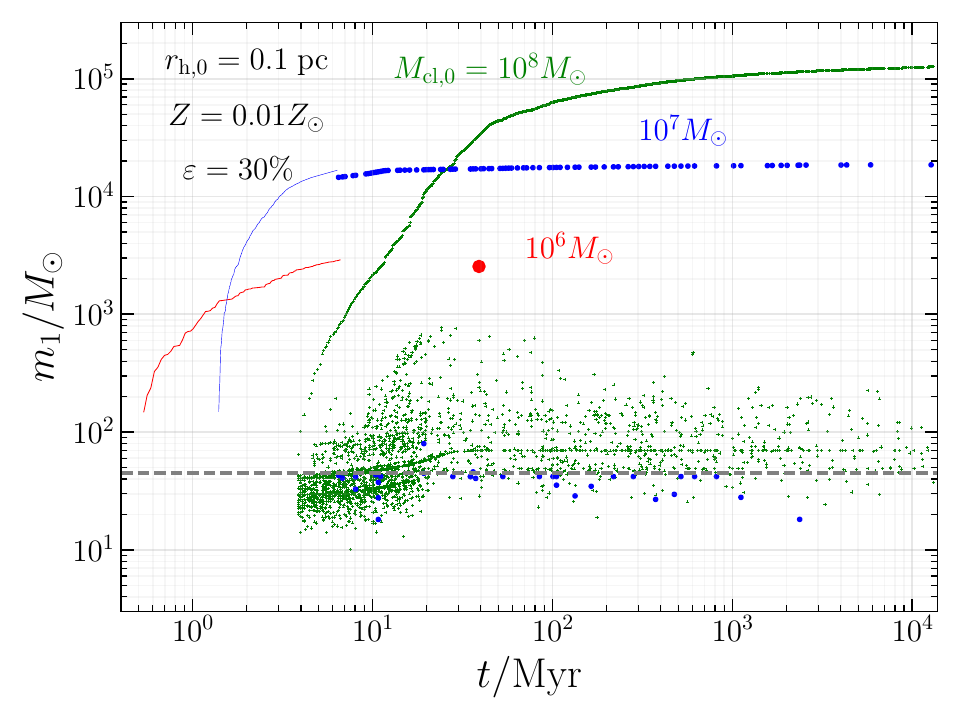}
    \caption{Merger tree history of three clusters with $r_{\rm h,0}=0.1~{\rm pc}$ and initial masses shown with text in the plot, 30\% of which fragmented into stars. The metallicity $Z$ was set to 1\% of the solar metallicity $Z_\odot=1.4\%$~\cite{2009ARA&A..47..481A}. Every point corresponds to a binary merger episode at time $t$ with primary mass $m_1$. Models with $M_{\rm cl}=10^6M_\odot$ and $M_{\rm cl}=10^7M_\odot$ experienced an initial phase of runaway stellar collisions, and the evolution of the supermassive star---which collapses into an IMBH at $t\simeq7$~Myr---is shown by the thin solid lines. The horizontal dashed line at $m_1\simeq45M_\odot$ represents the lower edge of the pair-instability supernova gap~\cite{Fryer:2011cx}.}
    \label{fig:tree_histories}
\end{figure}

As the number of retained stellar black holes (SBHs)---remnants of massive star evolution---can be up to hundreds of thousands, there can potentially be a large number of mergers among SBHs in a relatively small timescale.
The reaction rate between SBHs is enhanced by mass segregation (in an attempt to reach energy equipartition with low-mass stars, SBHs sink toward the center~\cite{Lee:1994nq}).
Figure~\ref{fig:tree_histories} shows the merger tree history of three NSC simulations, varying the initial total mass.

The two lighter simulated NSCs (red and blue points) collapse on a timescale $<3$~Myr and a head-start IMBH forms via stellar collisions.
Once the supermassive star collapses, the IMBH continues to grow slowly by merging with SBHs over the course of a Hubble time.
The runaway star, due to rejuvenation as it merges with smaller stars, can survive a little longer than 3~Myr.
Some gas accretion affects the IMBH of the $10^7M_\odot$ model in the first $\simeq15$~Myr.
Even though the $10^6M_\odot$ cluster initially had an escape velocity of a few 10~km/s---which decreases as it expands and evaporates---the IMBH is not ejected, because IMBH-SBH events are low-mass ratio inspirals whose merger products hardly receive a significant GW kick~\cite{Lousto:2009mf}. 
In our model, GW kicks are computed using the \textsc{Precession} code~\cite{Kesden:2014sla,Gerosa:2015tea,Gerosa:2016sys}.

\begin{figure}
    \centering
    \includegraphics[width=0.5\textwidth]{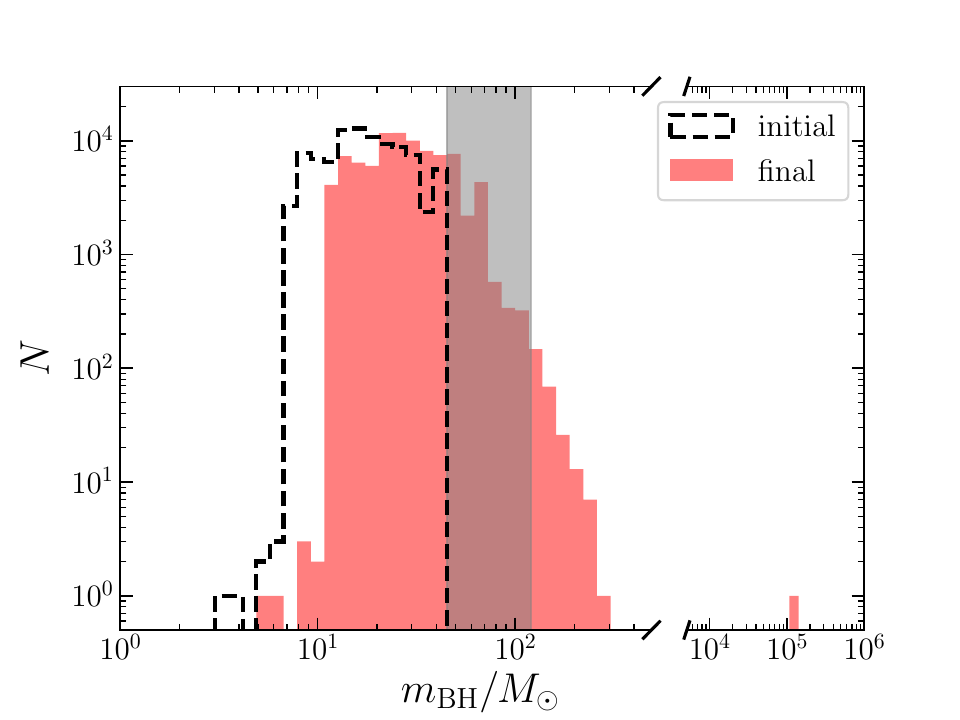}
    \caption{Histogram of BH masses when the BH subsystem forms (black dashed) and at the end of the simulation (red). The gray region denotes the limits of the PISN upper mass gap, and extends from $45M_\odot$ to $120M_\odot$.}
    \label{fig:gas_accretion_effect}
\end{figure}

\begin{figure*}
    \centering
    \includegraphics[width=\textwidth]{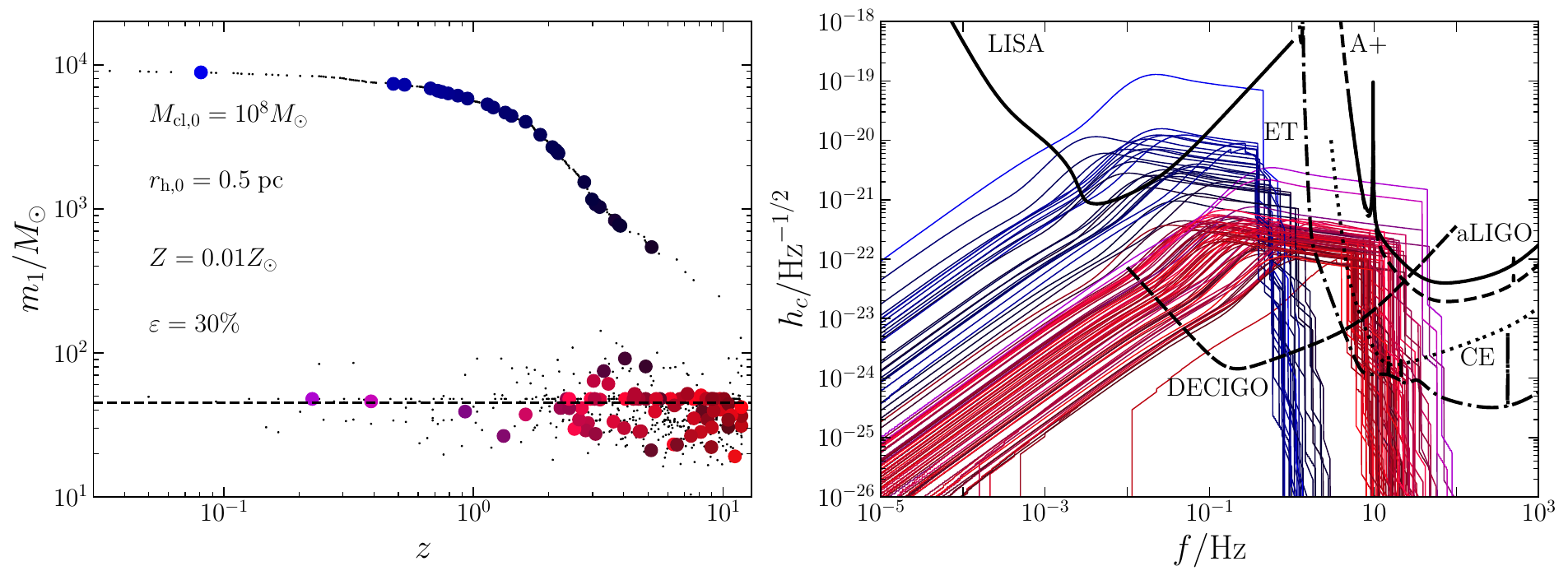}
    \caption{{Left panel:} merger tree history of a NSC (with initial conditions shown in the inset) and formation of an IMBH by repeated BH mergers. The points correspond to the primary masses $m_1$ of inspirals that merged at redshift $z$. A random sample of 100 mergers is shown with large colored dots. {Right panel:} the total characteristic GW strain $h_c$ of the 100 random events highlighted in the left panel, as a function of detector-frame GW frequency $f$.
    All BBHs were evolved adiabatically using Peters' formalism~\cite{Peters:1964zz} from their formation until the last stable orbit. The source distance is the luminosity distance of each merger at the redshift given by the x-ordinate value in the left panel.
    The black lines (solid, dashed, and dotted) correspond to $\sqrt{fS_n(f)}$ for future GW observatories. The sensitivity curves for LISA, DECIGO, and ground-based detectors (ET, CE, A+) were obtained from Refs.~\cite{Robson:2018ifk},~\cite{Yagi:2011wg}, and~\cite{Borhanian:2022czq}, respectively. The strain for aLIGO can be found online~\cite{AdLIGONoise}. The colors of the GW-strain lines in the right panel are mapped one to one with those in the left panel.}
    \label{fig:gw_history}
\end{figure*}

The heaviest cluster collapses on a timescale $>3$~Myr, and thus stellar collisions have no effect.
Regardless, the conditions in the core are so extreme that repeated BH mergers and gas accretion rapidly build up the mass of a $5\times10^4~M_\odot$ IMBH within 100~Myr.
Such a process has been referred to as a {\it cluster catastrophe}~\cite{Zevin:2022bfa}.

In fact, even SBHs experience gas accretion, increasing their mass up to hundreds of solar masses.
This can be observed in the thousands of mergers among SBHs that shift the BH pileup from $\simeq35M_\odot$ (a feature of the BH mass spectrum in low-metallicity environments~\cite{Belczynski:2016jno}) to $\simeq75M_\odot$, thus polluting the pair-instability supernova (PISN) gap~\cite{Woosley:2021xba}.
Figure~\ref{fig:gas_accretion_effect} shows the effect of gas accretion on BH masses. In particular, the lightest BHs grow by less than $4M_\odot$, while BHs in the pileup grow by $\sim30M_\odot$, as evidenced by the secondary peak that moves from $45M_\odot$ to $75M_\odot$. The distribution is also skewed due to hierarchical mergers: these mergers produce the high-mass tail that penetrates further into the PISN gap, extending up to $300M_\odot$.

After gas removal, the system becomes dry and the growth continues via mergers.
The growth is {\it dictatorial}~\cite{Kovetz:2018vly} in the sense that no second IMBH of comparable mass forms in the same environment, and any such attempt will fail as the primary IMBH will swallow it.
In this simulation, the massive BH reaches a mass of a little above $10^5M_\odot$ within a Hubble time.

\section{GWs from eccentric inspirals}
\label{sec:GWs_from_eccentric_inspirals}

The majority of the mergers in our simulations are GW captures between two BHs (one of which might be the IMBH), during which, a highly eccentric inspiral merges promptly~\cite{OLeary:2008myb}.
The right panel of Fig.~\ref{fig:gw_history} shows the characteristic GW strain amplitude $h_c$ (computed following Ref.~\cite{Bonetti:2020jku}) including the first ten harmonics, as calculated for a small random set of 100 inspirals from the simulation of the $10^8M_\odot$ cluster in the left panel. We neglect for simplicity the merger-ringdown part of the signal.

The value of the initial half-mass radius chosen for this simulation ($r_{\rm h,0}=0.5$~pc) is larger than the value chosen for Fig.~\ref{fig:tree_histories} in order to effectively switch off gas accretion and focus on growth driven by black hole mergers.
There are two broad categories of inspirals: SBH-SBH and SBH-IMBH events.

In the figure we also plot the sensitivity curves of future ground-based and space-borne GW observatories.
For consistency, we plot the noise amplitude defined as $\sqrt{fS_n(f)}$ \cite{Moore:2014lga}, where $S_n(f)$ is the noise power spectral density of each detector.
The area between the characteristic strain of each event and the noise amplitude is (very roughly) indicative of the signal-to-noise ratio of that event in a particular detector.

\begin{figure*}
    \centering
    \includegraphics[width=1\textwidth]{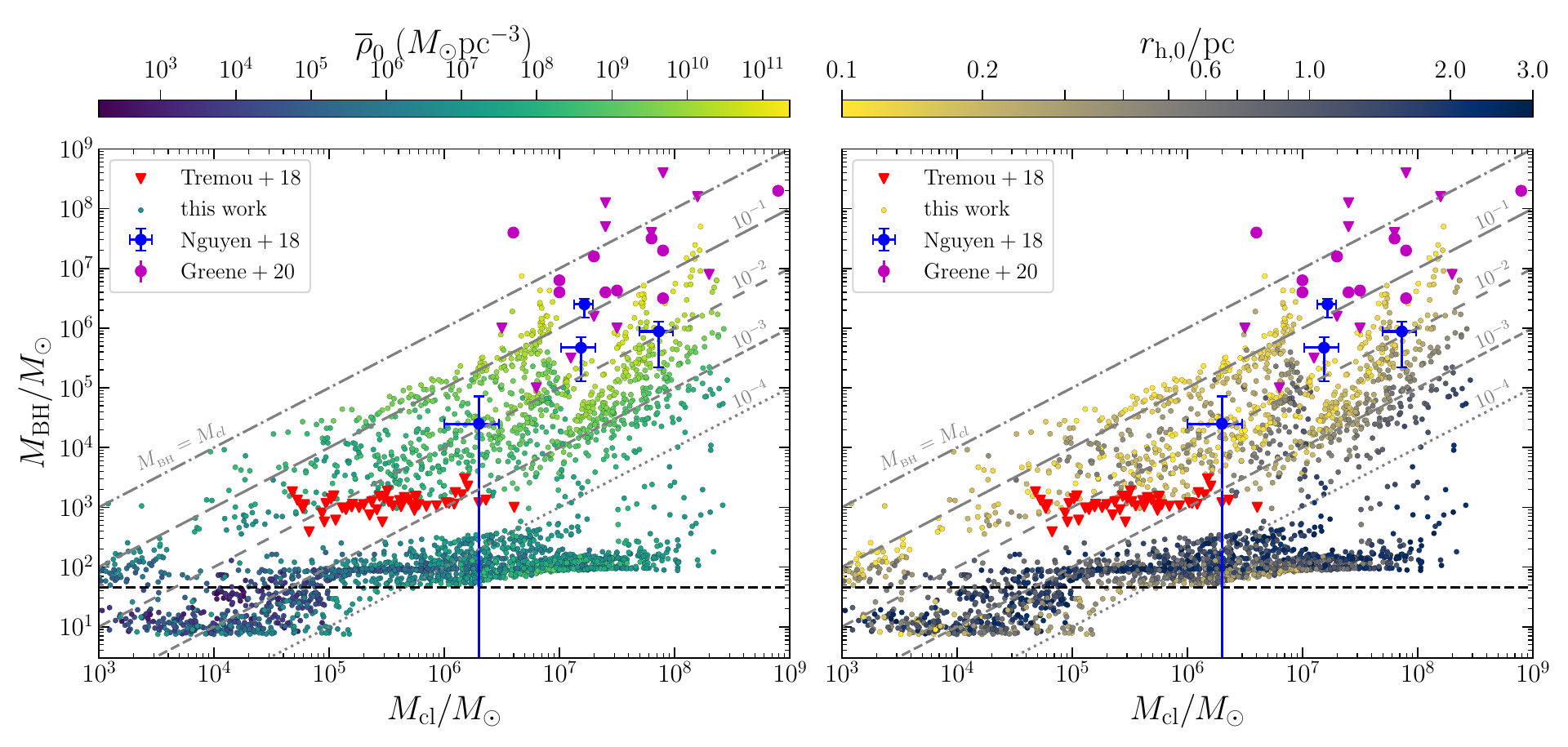}
    \caption{Mass of the heaviest BH formed versus the mass of the cluster at $z=0$ for a population of simulated NSCs. The color bar shows the initial mean density (left panel) and initial half-mass radius (right panel). The blue, magenta, and red points are observational data from Refs.~\cite{2019ApJ...872..104N}, \cite{2020ARA&A..58..257G}, and~\cite{Tremou:2018rvq}, respectively, while the magenta and red inverted triangles correspond to upper bounds in IMBH mass. The gray dashed lines show the fraction of the cluster's mass in the IMBH.}
    \label{fig:Mcl_MBH_plot}
\end{figure*}

\section{Probability of IMBH formation}
\label{sec:Probability_of_IMBH_formation}

We simulate thousands of NSC models with initial masses drawn log uniformly in the range $10^{4}$--$10^{9}M_\odot$, initial half-mass radii sampled in $0.1$--$3~{\rm pc}$, metallicity set to 0.01 of the solar metallicity, and for three discrete values of $\varepsilon=3\%,10\%,30\%$.
Each cluster is evolved until today ($z=0$), and we assume that fragmentation took place at $z=13$, which essentially corresponds to simulating our metal-poor NSCs for a Hubble time.

In Fig.~\ref{fig:Mcl_MBH_plot} we plot the heaviest BH mass as a function of the cluster mass, at the end of each simulation.
Every point represents a single NSC, and points are colored either by their initial mean density (left panel) or initial half-mass radius (right panel).
The initial mean density is defined as $\overline{\rho}_0\equiv3M_{\rm cl,0}/(4\pi r_{\rm h,0}^3)$.
We also plot observational data of NSCs along with their measured central BH masses (from Refs.~\cite{2019ApJ...872..104N,2020ARA&A..58..257G}).

\begin{figure}
    \centering
    \includegraphics[width=0.5\textwidth]{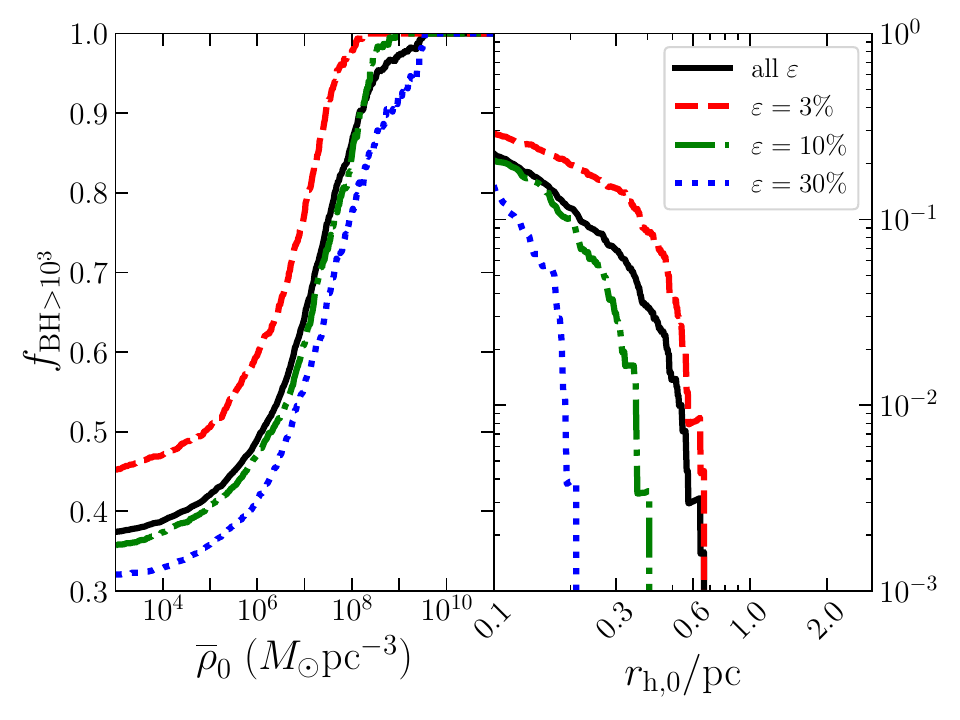}
    \caption{Fraction of clusters that formed a $>10^3M_\odot$ IMBH with $\overline{\rho}_0$ ($r_{\rm h,0}$) larger than the x-axis value in the left (right) panel, respectively. In the right panel, we considered only clusters with a present-day mass $<10^6M_\odot$. Colored lines refer to specific values of $\varepsilon$, while black solid lines account for all simulated NSCs (summed with a uniform weight).}
    \label{fig:fBH>1e3}
\end{figure}

The left panel of Fig.~\ref{fig:fBH>1e3} shows that in these simulations the formation of a $>10^3M_\odot$ IMBH is very likely (with a probability of $\approx60\%$) in clusters for which $\overline{\rho}_0>10^7M_\odot\,{\rm pc}^{-3}$,  while if $\overline{\rho}_0>10^{10}M_\odot\,{\rm pc}^{-3}$ IMBH formation is inevitable.
Furthermore, there is at least a 30\% chance that a sparser cluster ($\overline{\rho}_0<10^5M_\odot\,{\rm pc}^{-3}$) may harbor a low-mass IMBH, which typically forms via stellar collisions.

The existence of IMBHs in GCs has perplexed astronomers who tried to find them in their cores~\cite{2010ApJ...719L..60N,Gebhardt:2005cy,Gerssen:2002iq,2017IJMPD..2630021M}.
It is possible that the conditions in GCs are not favorable for the dynamical growth of IMBHs~\cite{Miller:2012ys,Gerosa:2019zmo} due to their small escape speeds (commonly a few 10~km/s), which limits the retention of IMBHs formed through a hierarchical BH runaway. Current constraints place the mass of heavy BHs to no more than $\approx10^3M_\odot$ in the Milky Way GCs~\cite{Tremou:2018rvq}.
We show these upper bounds in both panels of Fig.~\ref{fig:Mcl_MBH_plot} by the inverted red triangles.
Nevertheless, our simulations indicate the formation of some IMBHs in systems with properties that resemble those of GCs.
These typically correspond to compact clusters with $r_{\rm h,0}\ll1$~pc, which assembled their IMBHs via runaway stellar collisions.
In GCs, the IMBH cannot grow via repeated BH mergers, because the GW kick imparted to the growing BH often exceeds the escape velocity, and residual gas is removed too quickly to have an effect (see Fig.~\ref{fig:param_sp}).

To compare our results with the current constraints, we consider only clusters with present-day masses $<10^6M_\odot$.
For a given $r_{\rm h,0}$ we find only a fraction of all simulated systems to form an IMBH above $10^3M_\odot$ (no larger than about 30\%).
A cluster with $r_{\rm h,0}>0.5~{\rm pc}$ can form an IMBH that does not respect the upper bounds on GCs with probability $\simeq2\%$, which is consistent with the nonobservation of an IMBH in the 47 systems analyzed in Ref.~\cite{Tremou:2018rvq}.
This constraint weakens with $\varepsilon$, because the efficiency of stellar collisions depends on the relaxation timescale [cf.~Eqs.~\eqref{eq:half_mass_relaxation_timescale} and~\eqref{eq:collision_timescale}].

The absence of indisputable evidence for massive BHs in GCs allows us to constrain their initial half-mass radius to be not much smaller than 0.5~pc for efficiently ($\varepsilon\gtrsim10\%$) star forming GCs (see Fig.~\ref{fig:fBH>1e3}, right panel), otherwise there should be evidence of $\sim10^4M_\odot$ IMBHs in some GCs.
This conclusion is circumstantially supported by astronomical observations.
Young massive clusters ($M_{\rm cl}>10^4M_\odot$) are dense collisional environments detected in the local Universe with ages less than $100$~Myr~\cite{2010ARA&A..48..431P}, and thus have sizes and masses that are not very different from their initial conditions and not as evolved as GCs.
According to data (see, e.g., Fig.~9 in Ref.~\cite{2019ARA&A..57..227K}), they scarcely have half-mass radii less than $0.5$~pc.
If we accept that young clusters are progenitors of GCs, as some theories suggest~\cite{2014CQGra..31x4006K}, then it is very likely that the initial conditions of GCs were similar to present-day young massive clusters---albeit at lower metallicity.

\section{Conclusions}
\label{sec:Conclusions}

We studied the formation of IMBHs in NSCs evolved to the present epoch, and compared our results with local observations.
We found that the nonobservation of massive BHs in GCs suggests that it is unlikely that they formed with a half-mass radius $<0.6$~pc.
Recent work also found that $>10^3M_\odot$ IMBHs can form with high probability in large-mass, high-density star clusters~\cite{Trani:2022pbg}.

Our treatment of stellar mergers relies on the collision rate calibrated to $N$-body simulations of Ref.~\cite{PortegiesZwart:2002iks}. More recent $N$-body studies of IMBH growth from stellar mergers reported the formation of smaller IMBHs, with typical masses of order at most a few hundred $M_\odot$~\cite{Kremer:2020wtp,DiCarlo:2021att,Prieto:2022uot}. Note, however, that the smallest cluster simulated via $N$-body studies has $r_{\rm h,0}\sim 0.5$~pc~\cite{Kremer:2020wtp}. In our simulations, stellar collisions are likely to form $>10^3M_\odot$ IMBHs only in the most compact clusters with $r_{\rm h,0}\sim0.1$~pc. Even in that case, the probability of forming such IMBHs ranges between $\sim 13\%$ and $\sim 30\%$, depending on the star formation efficiency (see Fig.~\ref{fig:fBH>1e3}). Furthermore, very massive star runaways, whose evolution is highly uncertain, may lose mass by stellar winds at a rate faster than their growth through repeated stellar mergers, resulting in lower-mass remnants~\cite{Mapelli:2016vca,2009A&A...497..255G}. In summary, if the possibility of the formation of supermassive stars in ultracompact metal-poor clusters is realized in nature, it implies rather stringent constraints on the initial half-mass radius $r_{\rm h,0}$ of GCs.

Our assumption of a Plummer profile is somewhat conservative, because it has been shown that a cuspy distribution of stars may develop around the central IMBH~\cite{1976ApJ...209..214B,Alexander:2008tq}.
Cuspy profiles would enhance the interaction rates in the core and boost the growth of the IMBH seed.

IMBH seeds can further grow by accretion in the context of cosmological hierarchical assembly~\cite{Yoo:2004ze}.
If two merging galaxies host a central IMBH each, then an IMBH-IMBH binary may form once the two IMBHs approach each other.
A recent study showed that dynamical friction and binary-single hardening lead to the efficient merging of such IMBH-IMBH pairs in the cores of nucleated dwarf galaxies~\cite{Khan:2021jqf}.

Asymmetry studies show that some 4\%--5\% of dwarf galaxies have undergone major mergers~\cite{2014MNRAS.445.1157C,2022MNRAS.tmp.3118G}. Successive early merging events, when the dwarfs are gas rich and gas accretion also plays a role, should be sufficient 
to allow seeding of SMBH by IMBHs in massive galaxies. This remains to be quantitatively demonstrated via merger-tree analysis, a topic to which we intend to return. However, since most dwarf galaxies have not experienced mergers in the past,  IMBH seeds remain as relics to the present day without growing their masses beyond the range that we have predicted.

In this work, we followed the monolithic formation of NSCs by simplifying the star assembly history to a single starburst at high redshift. An alternative scenario for the formation of NSCs involves the buildup through the inspiral of GCs and their coalescence at the center~\cite{2023ApJ...944...79C}.
It is also likely that such a pathway influences the growth of the IMBH. Nevertheless, such a study goes beyond the scope of our current work.
While we acknowledge that the formation and evolution of star clusters---let alone that of NSCs---is a highly nontrivial process~\cite{2019ARA&A..57..227K,Longmore:2014epa}, we used a simplified picture for the purposes of this study.

Observations of gravitational runaways can be made with future GW observatories, such as LISA~\cite{LISA:2017pwj} and DECIGO~\cite{2017JPhCS.840a2010S}, which will observe mergers of IMBHs and SBHs out to high redshift.
In the local Universe, IMBHs that reside in the centers of dwarf galaxies may disrupt white dwarfs with an associated nuclear detonation transient~\cite{MacLeod:2015jma} which, together with detections of intermediate-mass ratio inspirals, can unambiguously disclose the existence of the elusive IMBHs.

Finally, we cannot resist noting that six young massive star clusters (with estimated stellar masses $10^{6}$--$10^{7}M_\odot$, found in the reionization epoch via lensing of the Sunrise arc~\cite{2022arXiv221109839V}) are unusually compact (down to $\sim 1$~pc) and potential candidates for early NSC formation according to our prescription for IMBH factories.

\acknowledgments
We thank Andrea Antonelli, Floor Broekgaarden, Roberto Cotesta, Mark Cheung, Giacomo Fragione, Ken Ng, Luca Reali, Vladimir Strokov, and Silvia Toonen for discussions and comments.
K.K. and E.B. are supported by NSF Grants No. AST-2006538, PHY-2207502, PHY-090003 and PHY20043, and NASA Grants No. 19-ATP19-0051, No. 20-LPS20-0011 and No. 21-ATP21-0010.
This work was carried out at the Advanced Research Computing at Hopkins (ARCH) core facility (\url{rockfish.jhu.edu}), which is supported by the NSF Grant No. OAC-1920103.
The authors acknowledge the Texas Advanced Computing Center (TACC) at The University of Texas at Austin for providing {HPC, visualization, database, or grid} resources that have contributed to the research results reported within this paper~\cite{10.1145/3311790.3396656,tacc}.

\appendix

\section*{Appendix: Heating efficiency of three-body BH binaries in NSCs}

Reference~\cite{Miller:2012ys} determined that binary star heating would become inefficient in systems with velocity dispersion $\sigma$ that exceeds about 40~km/s.
The limit comes from the fact that the hard-soft boundary corresponds to a value that is smaller than twice the solar radius. Therefore there cannot be hard binary stars to provide energy through binary-single interactions in sufficiently ``hot'' environments.
Their analysis, however, did not consider stellar BHs, which are extremely compact and can form hard pairs in environments with velocity dispersion that well exceeds 40~km/s.
Hence, BH binaries can provide a source of energy for the cluster, with a mechanism that is described in Ref.~\cite{Breen:2013vla}.

The energy produced via binary-single interactions in the core of the BH subsystem is then transferred to the ambient stars via relaxation, the whole system is balanced, and collapse is prevented.
Nevertheless, at sufficiently high values of $\sigma$ the hard-soft boundary of BH binaries is pushed down to very small values, and the GW emission rate is high enough that binaries merge before they heat up the system through interactions. In that case the heating efficiency drops.
Since the merger time depends on eccentricity, the heating efficiency is a number that ranges from zero to one, because a fraction of three-body binaries---if eccentric enough---merges promptly, while the rest can heat up the cluster.

\begin{figure}
    \centering
    \includegraphics[width=0.5\textwidth]{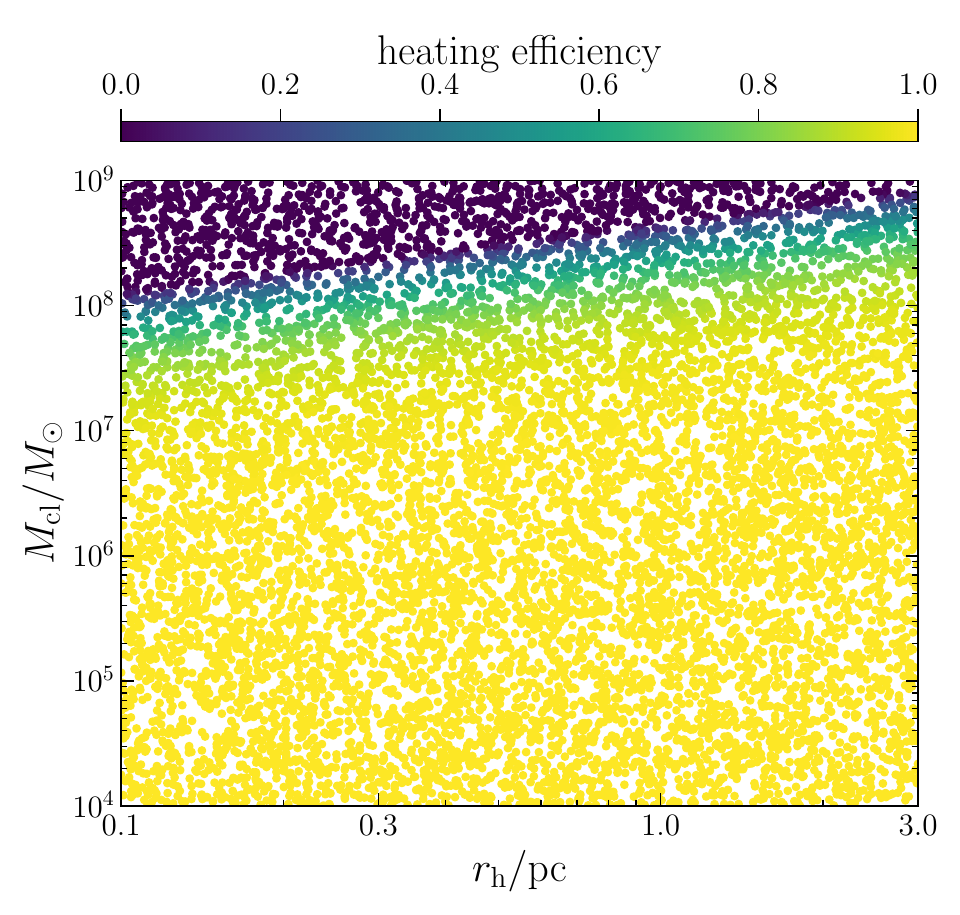}
    \caption{Star cluster mass vs. half-mass radius parameter space. The color bar corresponds to the fraction of three-body binaries that are able to generate energy via binary-single interactions, assuming a thermal eccentricity distribution.}
    \label{fig:heating_efficiency}
\end{figure}

Under the assumption of a thermal eccentricity distribution, we have computed the heating efficiency of three-body binaries in the parameter space $(r_{\rm h}, M_{\rm cl})$.
For this, we have randomly sampled $10^4$ star clusters with $r_{\rm h}\in[0.1,3.0]~\rm pc$ and $M_{\rm cl}\in[10^4, 10^9]M_\odot$ while to compute the heating efficiency we averaged over $10^4$ realizations of binary eccentricity per grid point. 
In particular, we have compared the GW coalescence time with the interaction time. If the former is larger than the latter, then the three-body binary is able to generate entropy in the cluster.
We show our results in Fig.~\ref{fig:heating_efficiency}.
We find that for star clusters with $M_{\rm cl}<10^8M_\odot$ and $r_{\rm h}>0.1~\rm pc$, three-body binaries are able to provide the energy required to support the core of the cluster from collapsing, and the balanced evolution of Ref.~\cite{Breen:2013vla} is valid.
This is most of the parameter space we have simulated (see Fig.~\ref{fig:Mcl_MBH_plot}), and this leads to the formation of massive BHs.

To carry out our calculations we have made the following assumptions for the BH subsystem.
The stellar velocity dispersion is $v_{\rm star}=(0.4GM_{\rm cl}/r_{\rm h})^{1/2}$;
the mean stellar mass is $0.6M_\odot$;
the mean BH mass is $20M_\odot$ (in low-metallicity environments);
the BH velocity is $v_{\rm BH}=[m_{\rm star}/(0.7m_{\rm BH})]^{1/2}v_{\rm star}$ (see~\cite{Lee:1994nq});
the BH hard-soft boundary is $a_{\rm BH}=Gm_{\rm BH}/(4v_{\rm BH}^2)$ (see~\cite{Sesana:2006xw});
the number of BHs is $N_{\rm BH}=3\times10^3[M_{\rm cl}/(10^6M_\odot)]$ (see~\cite{Fragione:2023kqv});
the BH half-mass radius is $r_{\rm h,BH}=0.7r_{\rm h}N_{\rm BH}m_{\rm BH}^2/(m_{\rm star}M_{\rm cl})$ (see~\cite{Lee:1994nq});
the BH number density is $n_{\rm BH}=N_{\rm BH}/(4r_{\rm h,BH}^3)$, and
the eccentricity of BH binaries is assumed to follow a thermal distribution.

\bibliography{refs2}

\end{document}